\documentclass[twocolumn,showpacs,preprintnumbers,amsmath,amssymb,prb,aps]{revtex4-1}
\usepackage{epsfig}
\usepackage{epstopdf}
\usepackage{graphicx}
\usepackage{dcolumn}
\usepackage{bm}
\usepackage{textcomp}
\usepackage{array}
\usepackage{subcaption}
\usepackage{booktabs}

\begin{document}

\title{Investigating the effect of temperature dependent many-body interactions on bulk electronic structures and the robust nature of (001) surface states of SnTe}

\author{Antik Sihi$^{1,}$}
\altaffiliation{sihiantik10@gmail.com}
\author{Sudhir K. Pandey$^{2,}$}
\altaffiliation{sudhir@iitmandi.ac.in}
\affiliation{$^{1}$School of Basic Science, Indian Institute of Technology Mandi, Kamand - 175075, India\\
$^{2}$School of Engineering, Indian Institute of Technology Mandi, Kamand - 175075, India}

\date{\today}

\begin{abstract}

  Recently, SnTe has gained attention due to its non-trivial topological nature and eco-friendly thermoelectric applications. We report a detailed temperature dependent electronic structure and thermodynamic properties of this compound using DFT and GW methods. The calculated values of bandgaps by using PBEsol and $G_0W_0$ methods are found to be in good agreement with the experiment, whereas mBJ underestimates the bandgap. The estimated value of fully screened Coulomb interaction ($W$) for Sn (Te) 5$p$ orbitals is $\sim$1.39 ($\sim$1.70) eV. The nature of frequency dependent $W$ reveals that the correlation strength of this compound is relatively weaker and hence the excited electronic state can be properly studied by full-$GW$ many-body technique. The plasmon excitation is found to be important in understanding this frequency dependent $W$. In order to describe the experimental phonon modes, the long range Coulomb forces using nonanalytical term correction is considered. The temperature dependent electron-electron interactions (EEI) reduces the bandgaps with increasing temperature. The value of bandgap at 300 K is obtained to be $\sim$161 meV. The temperature dependent lifetimes of electronic state along W-L-$\Gamma$ direction are also estimated. This work suggests that EEI is important to explain the high temperature transport behaviour of SnTe. We have also explored the possibility of protecting the (001) surface states via mirror and time-reversal symmetry. These surface states are expected to be robust against the point and line defects. 
    
\end{abstract}

\maketitle

\section{Introduction} 
  
  Topological crystalline insulator (TCI) is a new class of material within the realm of topological quantum systems, where crystal symmetry protects the metallic surface states \cite{fuprl,bansil}. Recently, SnTe is marked as the first TCI member after both theoretical \cite{hsieh} and experimental \cite{tanaka} observations. This material is also a promising candidate for eco-friendly thermoelectric material with narrow semiconducting bandgap \cite{abanik}. In order to predict the bandgap value using theoretical approach, the first principle based density functional theory (DFT) is widely used with local-density approximation (LDA) or generalized gradient approximation (GGA) exchange-correlation (XC) functional. In general for semiconducting materials, DFT underestimates the value of experimental bandgap. But, the predicted bandgap using mBJ exchange potential within DFT method provides good agreement with the experimental observation for different class of materials \cite{koller2011,koller2012}. However, the accuracy of this method for finding the bandgap depends on the proper choice of parameters, which may be varied on changing of materials \cite{koller2012,hjiang2013}. This seems to be a tedious and nonphysical situation for any newly predicted material. Therefore, the parameter free Hedin's $GW$ approximation (GWA), which is based on many-body perturbation theory with consideration of electrons screening, comes to tackle this problem \cite{lhedin}. When the Dyson equation within this GWA is solved by using single iteration of the one particle Green's function ($G$) and the fully screened Coulomb interaction ($W$), then this $ab$ $initio$ method is called as $G_0W_0$ \cite{martin}. In recent years, to get the improved ground state electronic structure for different materials, the $G_0W_0$ method is widely used due to its lower computational cost. It is noted that only few works on SnTe are found for comparing the bandgap between experiment and $G_0W_0$ method \cite{eremeev,aguado}. However, it is interesting to note that the calculated bandgap values obtained from mBJ are nicely matched both with experiment and $G_0W_0$ method for different materials as shown by Koller $et$ $al$ \cite{koller2011,koller2012}. It is also mentioned that the computational cost of mBJ calculation is very less than $G_0W_0$ method. Hence, it will be interesting to carry out a comparative study for calculating the bandgap using all these mentioned methods for this topological material. This study will establish the more strength of individual method for finding the bandgap. In addition to this, it is known that the effect of electronic correlations gives the exciting new physics for different materials.

  The strength of electronic correlation effect of any material can be estimated from the frequency ($\omega$) dependent $W$ \cite{antik}, which also provides the information about plasmon excitation \cite{arya2006, miyake2009, sakuma2013, amadon2014}. Nowadays, different many-body techniques ($i.e.$ fully self-consistent $GW$ (full-$GW$), DFT + dynamical mean field theory (DMFT), $GW$ + DMFT etc.) are developed to study different class of compounds. The strength of correlation suggests the proper choice of many-body method for studying different physical properties of the materials. But to the best of our knowledge, this aspect is not explored for SnTe, where some insightful information about strength of correlation effect can be obtained for this compound. In order to study this correlation effect due to electron-electron interactions (EEI), full-$GW$ is generally suggested for the s and/or p-block elements \cite{martin}. This many-body interactions effect always brings more deeper understanding of any material, when temperature dependency is also included.

   Recently, the effect of temperature on electronic structure of this material due to electron-phonon interactions (EPI) is studied by Querales-Flores \textit{et} \textit{al} \cite{flores}. But, the effect of EEI due to changing temperature is still not investigated. This study may provide the importance of the excited electronic state for this non-trivial topological material, which have 5$p$ as the outer most orbitals both for Sn and Te atoms. Thus, full-$GW$ method is expected to be the perfect candidate for this study. In this method, both fully self-consistent $G$ and $W$ are used to solve the Dyson equation based on all-electron many-body perturbation theory \cite{martin}. This method satisfies the conservation laws of momentum, energy and particle number, where $G_0W_0$ violates all these conservations \cite{baym,baym2,dahlen}. The electronic temperature is introduced by calculating the $GW$ self-energy ($\Sigma$(\textbf{k},$\omega$) = $iGW$) within Matsubara-time domain \cite{chu-elk}. This finite temperature full-$GW$ technique is capable to successfully predict not only the spectral function for a fixed k-point, but also the temperature dependent bandgaps of many semiconducting materials \cite{chu-elk,faleev}. Also, the important information about the quasiparticle excitation can be possible to extract from this $\Sigma$(\textbf{k},$\omega$) \cite{martin}. This technique also has ability to capture the satellite features, which is useful to get the better insight of collective quasiparticle excitations \cite{martin}. It is known that all these methods ($i.e.$ DFT and GWA) give the information about the electrons. But, the materials formed with electrons and ions. These ions are seated on lattice points to form the crystal structure. The dynamics of these ions can be possible to understood with the help of phonon modes.

   It is found from theoretical observation that the topological phase of SnTe can be tuned via lattice deformations using dynamical phonon modes or strains \cite{kim}. Thus understanding the phonon modes of this compound is really important. Experimental phonon dispersion curve of this compound shows the degeneracy lifting of the optical phonon modes into longitudinal and transverse parts at $\Gamma$-point \cite{oneill}. Generally in ionic crystals, this degeneracy lifting can be obtained after considering long range Coulomb forces into theoretical calculation \cite{togo_nac}. It is known that SnTe is not a pure ionic crystal. Therefore in order to understand this degeneracy lifting, the effect of long range Coulomb forces on this compound's phonon modes need to be explored through nonanalytical term correction (NAC). These phonon modes are the key concept to understand the dynamical stability and thermodynamic properties of any material \cite{togo,shastri2018,shastri2019}. In addition to all these bulk properties, it will be interesting to observe the robustness of metallic surface states for this non-trivial topological compound.

  The metallic surface states of this compound form gapless Dirac cones for (111) and (001) surfaces, which are normal to (1-10) mirror plane \cite{liu,ytanaka,jwang}. The metallic surface bands for (111) plane of this compound are found both at $\bar{\Gamma}$- and $\bar{X}$-points \cite{liu,ytanaka}. However, in case of (001) surface, these bands are observed on the mirror symmetric line of $\bar{\Gamma}$-$\bar{X}$ \cite{liu,ytanaka87}. Therefore, the Dirac cones found from (111) surface are protected by mirror symmetry and time-reversal symmetry (TRS), whereas for (001) surface former one only protects these surface bands. Thus, it opens to think a new idea for (001) surface to find out the Dirac cone at $\bar{X}$ or $\bar{\Gamma}$-point. When this is achieved, then this Dirac cone will be protected by the mirror symmetry and TRS. Although in recent theoretical study, it is found that the non-trivial topological phase of this compound can be changed to trivial via induced pressure \cite{aguado}. Therefore, it is also important to observe the robustness of these metallic surface states due to different types of defects, which is necessary for device making application.

  In present work, we focus on SnTe for searching the suitable ground state electronic structure method, estimating the strength of correlation effect, importance of long rang Coulomb interaction for phonon calculation, effect of EEI due to changing the temperature and obtaining the Dirac cone at time-reversal invariant momenta point for (001) surface. The bandgap values obtained from PBEsol and $G_0W_0$ are nicely matched with experiment, whereas mBJ is underestimating. The calculated value of $W$ for Sn (Te) 5$p$ orbitals is $\sim$1.39 ($\sim$1.70) eV. After observing the behaviour of $W(\omega)$, full-$GW$ is chosen as suitable method for tackling the correlation effect of this compound. It is noted that the plasmon excitation helps to understand this $W(\omega)$. Here, the long range Coulomb forces are found to be important for good matching of phonon dispersion with experiment, which is implemented through NAC. The values of bandgap obtained from temperature dependent electronic structure study are showing monotonically decreasing nature with increasing temperature. The bandgap value at 300 K is found to be $\sim$161 meV, which is nicely agreed with experimental value. The EEI are seen to be more important than EPI, which is suggesting that EEI can not be ignored in understanding the high temperature transport behaviour of this compound. The imaginary part of self energy ($Im \Sigma (\textbf{k},\omega)$) for 300 K is estimated along W-L-$\Gamma$ direction. In addition to all, 14 layers calculation provides the Dirac cone at $\bar{X}$-point on (001) surface and resulting surface states seem to be robust against different types of defects.

\begin{figure*}
  \begin{center}
    \includegraphics[width=0.9\linewidth, height=9.5cm]{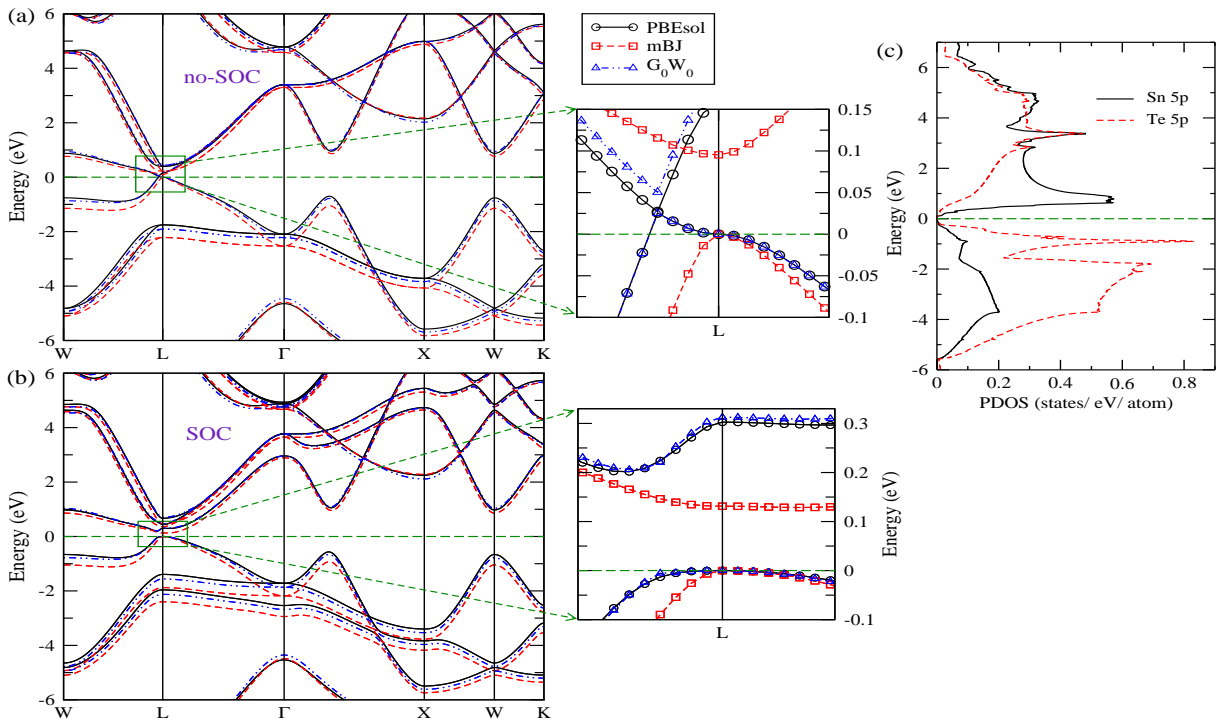} 
    \caption{(Colour online) Electronic band structure of SnTe (a) without and (b) with including SOC using PBEsol (black solid line with circle), mBJ (red dashed line with square) and $G_0W_0$ (blue dotted dash line with triangle). (c) Partial density of states (PDOS) of Sn 5$p$ and Te 5$p$ orbitals calculated using PBEsol.}
    \label{fig:}
  \end{center}
\end{figure*}

\section{Computational details}

  In this work, the full-potential linearized-augmented plane-wave (FP-LAPW) based WIEN2k code \cite{wien2k} is used for spin-unpolarized electronic structure calculation of SnTe with PBEsol XC functional \cite{pbesol} and mBJ exchange potential with GGA correlation \cite{mbj}. SnTe has the space group of $F$m-3m. The value of optimized lattice parameter using PBEsol functional is found to be 6.291 \AA, which is utilized for this study. This calculated value is nicely matched with experimental lattice parameter of SnTe for this space group\cite{bis, littlewood, dzhang}. 10\texttimes 10\texttimes 10 \textbf{k}-mesh size is chosen for this calculation. The Sn and Te atoms are occupied at Wyckoff positions (0,0,0) and ($\frac{1}{2}$,$\frac{1}{2}$,$\frac{1}{2}$), respectively. The muffin-tin radius for both Sn and Te atoms are set to be 2.5 bohr with the convergence criteria of 10$^{-4}$ Ry/cell for total energy calculation. $R_{mt} * K_{max}$ is fixed at 8.5 in entire calculation for obtaining the better convergence. The Born effective charge (BEC) of Sn (Te) atom is calculated by displacing any one of four Sn (Te) atoms along z-axis with amount of 0.002 in fractional coordinates. It is obtained with the force convergence criteria of 0.002 mRy/Bohr using BerryPi \cite{berrypi} code, which is included into the recent version of WIEN2k \cite{wien2k}. In order to obtain different Coulomb interaction using random-phase approximation (RPA) with Wannier basis function, GAP2 code is used \cite{jiang1, jiang2}. $G_0W_0$ calculation has been also carried out using this code. The phonon calculations for this compound are done by PHONOPY \cite{togo} code using finite displacement method via constructing 2\texttimes 2\texttimes 2 supercell. However, for the better accuracy of result, 21\texttimes 21\texttimes 21 mesh size is taken for calculating the phonon dispersions, phonon density of states (DOS) and thermal properties of SnTe. The full-$GW$ calculation in Matsubara time-domain is performed using Elk code \cite{elk}. In this calculation, 4\texttimes 4\texttimes 4 \textbf {q}-mesh size is employed along with 8\texttimes 8\texttimes 8 \textbf{k}-mesh. To transform the spectral function from imaginary to real axis, Pa$\acute{d}$e approximation \cite{pade} is chosen due to the lower computational cost and its' simple implementation. In order to calculate the surface states, one layer of SnTe (001) surface is defined by constructing the supercell of dimension 1\texttimes 1\texttimes 1 with 8 atoms, which is not usual for a typical surface state calculation. This one layer is repeated 14 times to make 14 layers with 128 atoms and 30 Bohr vacuum is provided along the $k_z$ direction to reduce the interaction between two consecutive layer. Spin-orbit coupling (SOC) is included through out this study except for phonon and full-$GW$ calculations.

\section{Results and Discussion} 

\subsection{Ground state electronic structure and correlation effect}

  In order to compare the ground state electronic structure of SnTe, the dispersion curves obtained from PBEsol, mBJ and $G_0W_0$ calculations are shown in Fig. 1(a) along the high-symmetric k-direction within the energy of -6.0 eV to 6.0 eV. These calculations are performed without including SOC. In case of valence band (VB), this comparison shows that the $G_0W_0$ bands nicely follow the PBEsol bands with a small shift along the entire k-direction. Moreover, it is also found that the maximum shifting of $G_0W_0$ bands with respect to PBEsol bands are observed along L-$\Gamma$ direction at $\sim$2.0 eV. Whereas, the mBJ bands move 0.2 - 0.5 eV towards the lower energy than the PBEsol bands and getting the maximum shift along L-$\Gamma$ direction. However, the calculated conduction bands (CB) from these three calculations are not showing any significant difference with each other. Furthermore, to get a better insight of the low-energy bands which are mainly responsible for topological and transport properties, the band structure obtained from these three calculations are also separately plotted around the L-point in Fig. 1(a). It is clearly seen from this figure that the PBEsol and $G_0W_0$ calculations are not showing the semiconducting ground state for SnTe. But, at least, this behaviour is observed from mBJ calculation. In this case, the bandgap value is found to be $\sim$95 meV, which underestimates the experimental value\cite{esaki,dimmock,tsu}. It is known that both Sn and Te atoms are comparatively heavier element which means that the inclusion of SOC in these calculations may provide the better electronic structure for this compound. Therefore, SOC is included in all these calculations and computed band structure is shown in Fig. 1(b). Here, it is important to note that the SOC correction in $G_0W_0$ calculation is considered at the DFT level \cite{jiang1, jiang2}. In overall, the figure shows the similar kind of comparative behaviour as obtained from Fig. 1(a) except observing some degeneracy lifting around the high-symmetric k-direction, which is a common nature observed after including SOC. But, around the $E_F$, the band structure is significantly changed as compared to earlier discussion, which leads to get the semiconducting ground state from all these calculations. It is also observed from the figure that the fundamental bandgaps obtained from PBEsol and $G_0W_0$ are indirect in nature, whereas mBJ shows the direct bandgap at L-point. Therefore after including SOC, the modified values of bandgap obtained using PBEsol, mBJ and $G_0W_0$ calculations are $\sim$0.20 eV, $\sim$0.13 eV and $\sim$0.21 eV, respectively. At this point, it is noted that the values of bandgap obtained from both of PBEsol and $G_0W_0$ calculations with SOC are nicely matching with the experimental value \cite{esaki,dimmock,tsu} of SnTe in cubic phase. However, mBJ is still underestimating the bandgap as compared to experimental value. Koller $et$ $al.$ \cite{koller2011} shows that for some compounds, the mBJ is not providing the proper bandgap when compared with the experimental observations. So, they improved this method with reparametrization of some coefficients and claimed to get the best matching of bandgap with experiment \cite{koller2012}. But, this scenario again opens the same problem for different compound and suggests to further improve this method for better accuracy with less system dependency. Moreover, it is clearly found from the figure that the calculated values of direct bandgap at the L-point from PBEsol and $G_0W_0$ are $\sim$0.30 and $\sim$0.31 eV, respectively. In case of direct bandgap, these calculated values are in good agreement with the experimentally observed optical absorption edge \cite{bylander,burke}, where optical gap is defined as the smallest direct gap between VB and CB at room temperature. At this point, one should keep in mind that Bylander $et$ $al.$ \cite{bylander} proposed the possibility of smaller indirect bandgap for SnTe, which is also evident from this calculation. Importantly, it is found that the curvature of all the bands obtained from both PBEsol and $G_0W_0$ are quite similar in nature. This behaviour suggests that the correction of electronic structure with considering EEI is not expected to affect the physical properties of this compound related with ground state electronic structure closer to the Fermi level. But, it shows significant changes in the electronic band structure specially for VB in lower energy region, which may provide the proper ground state electronic structure of this compound. Now, the partial DOS (PDOS) of 5$p$ orbitals for Sn and Te atoms using PBEsol is shown in Fig. 1(c) without including SOC, because it is not expected to change significantly after including SOC. It is observed from this figure that these bands in the energy window of -6.0 eV to 7.0 eV are mainly formed by the contribution of Sn 5$p$ and Te 5$p$ orbitals. Therefore, it will be very interesting to study different Coulomb interactions for these two orbitals which will provide the importance of the correlation effect.

 \begin{figure}
  \begin{center}
    \includegraphics[width=0.8\linewidth, height=5.6cm]{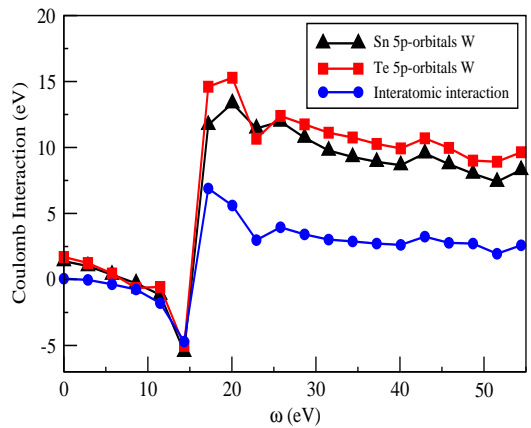} 
    \caption{(Colour online) Coulomb interaction as a function of $\omega$.}
    \label{fig:}
  \end{center}
\end{figure}

\begin{figure*}
  \begin{center}
    \includegraphics[width=0.8\linewidth, height=9.5cm]{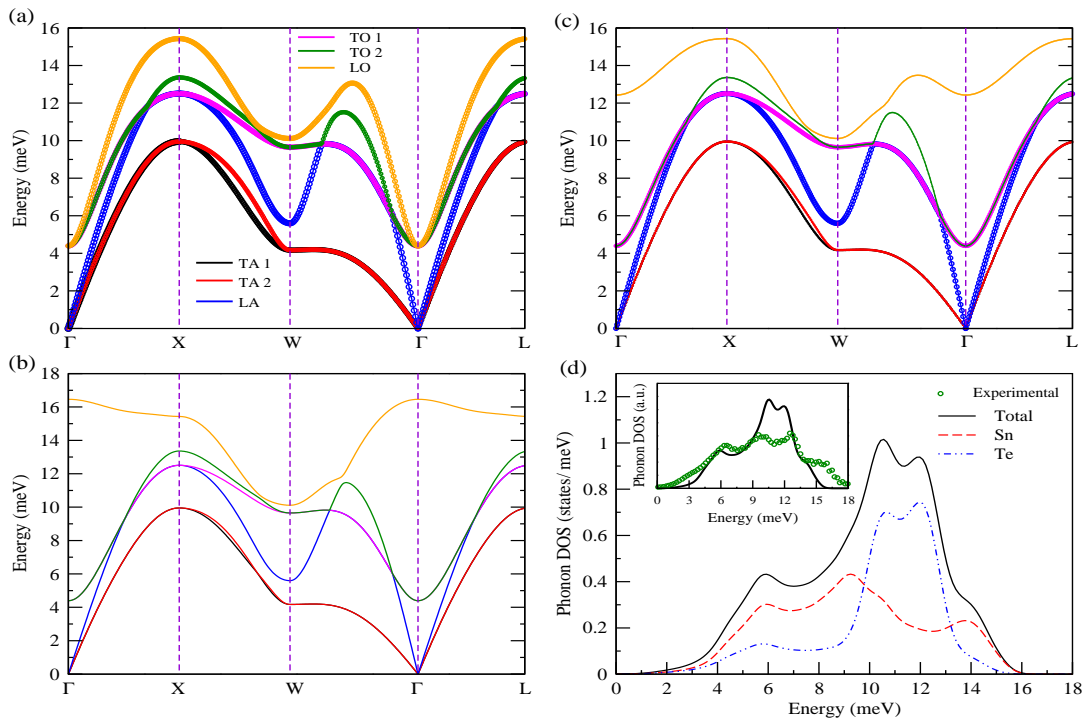} 
    \caption{(Colour online) Phonon band structure (a) not including nonanalytical term correction (NAC), including NAC with the value of Born effective charge (BEC) (b) $\sim$7.5 ($\sim$-7.5) and (c) $\sim$5.5 ($\sim$-5.5) for Sn (Te) atom. (d) Phonon TDOS (black solid line), phonon PDOS of Sn (red dashed line) and Te (blue dotted dash line) atoms. Inset shows the comparison between TDOS and experimental spectra\cite{li}.}
    \label{fig:}
  \end{center}
\end{figure*}

  Here, the bands obtained from -6.0 eV to 6.0 eV are selected for calculating the different Coulomb interactions using Wannier function. The calculated values of diagonal bare Coulomb interaction, full bare Coulomb interaction and on-site bare exchange interaction for Sn (Te) 5$p$ orbitals are $\sim$7.92 ($\sim$8.96) eV, $\sim$8.39 ($\sim$9.63) eV and $\sim$0.56 ($\sim$0.57) eV, respectively. The computed value of intra bare Coulomb interaction for Sn (Te) 5$p$ orbitals is $\sim$7.67 ($\sim$8.62) eV. In overall, the bare inter-atomic Coulomb interaction between Sn 5$p$ and Te 5$p$ orbitals is found to be $\sim$2.69 eV with a very small variation between diagonal and off-diagonal matrix elements. Here, the computed value of $W$ for Sn (Te) 5$p$ orbitals is $\sim$1.39 ($\sim$1.70) eV. The calculated values of inter-atomic Coulomb interaction Sn 5$p_i$ - Te 5$p_j$ (where, $i=x,y,z$ and $j=x,y,z$) are found to be $\sim$59 meV (when $i=j$) and $\sim$62 meV (when $i\neq j$), respectively. Therefore, the averaged value of inter-atomic Coulomb interaction is obtained as $\sim$61 meV. It is important to note that this number represents the strength of Coulomb interaction between two nearest neighbour site occupied by Sn 5$p_i$ and Te 5$p_j$ atoms in presence of electrons' screening due to other orbitals except these two.

  The effect of orbital screening on the electronic structure of SnTe can be understood from the plots of $\omega$ dependent $W$ for Sn 5$p$ and Te 5$p$ orbitals, as shown in Fig. 2. This figure also shows the $\omega$ dependent inter-atomic Coulomb interaction of this compound. Here, the value of $W$ for Te 5$p$ orbitals is slightly higher than Sn 5$p$ orbitals for the studied $\omega$. The minimum value obtained at $\sim$14.3 eV for Sn (Te) 5$p$ orbitals is $\sim$-5.5 ($\sim$-5.0) eV. This minimum peak may be due to a collective plasma oscillation, which is known as plasmon, because of internal transition of Sn (Te) $5p$ orbitals. In analogy with some previous work on different compounds, the similar kind of peak is observed at some certain energy and described the reason with the help of plasmon excitation \cite{arya2006,miyake2009,sakuma2013,amadon2014}. It is found that this peak's position is nicely agreed with the width of Sn (Te) $5p$ orbitals, which is also evident from Fig. 1(c). However, it is clearly observed from the figure that above $\omega$ = 14.5 eV, the values of $W$ for Sn 5$p$ and Te 5$p$ orbitals are rapidly increasing. This behaviour represents that the effect of plasmon is insignificant after this $\omega$. In the higher $\omega$ region, these values move to the values of their corresponding bare Coulomb interaction. Therefore from this scenario, it is understood that the orbital screening effect is insignificant for higher $\omega$ region but not in case of lower $\omega$ due to the major contribution of plasmon. The similar behaviour is also observed from the figure of $\omega$ dependent inter-atomic Coulomb interaction. The monotonic decrement of $W$($\omega$) till plasmon active region with weakly $\omega$ dependent suggests that the system is weakly correlated \cite{arya2004,miyake2008}. Thus, the full-$GW$ method based on many-body perturbation theory is expected to provide the good explanation of the excited electronic states for SnTe.

\subsection{Phonon and thermodynamic properties}

  Now, the calculated phonon band structure along high-symmetric k-direction is plotted in Fig. 3(a) for investigating the phonon and thermal properties of this compound. In order to find whether the effect of long range Coulomb forces on the vibrational properties of this compound is significant or not, firstly the calculation is carried out without NAC \cite{nac}. It is clear from the figure that no negative frequency has been observed for this compound suggesting the mechanical stability in this structure. Here, it is also observed from the figure that six phonon branches are found, which is expected for this compound. These phonon bands are further categorizes into two transverse acoustic (TA), one longitudinal acoustic (LA), two transverse optical (TO) and one longitudinal optical (LO) branches, which is mentioned in the figure. The two TA and one LA branches are degenerate at $\Gamma$-point. But, this degeneracy is lifted along the observed high-symmetric k-direction except at $\Gamma$-point. However, these two TA branches keep their degeneracy along $\Gamma$-X and W-$\Gamma$-L directions, whereas the degeneracy lifting is seen along X-W k-path. The LA branch touches one TO branch with containing same phonon energy around X \& L-points and along W-$\Gamma$ direction. In case of optical branches, they are all degenerate at $\Gamma$-point with energy $\sim$4.6 meV. But, O'Neill $et$ $al.$ \cite{oneill} show in their experimental work that only two TO branches are degenerate at this k-point. Thus, this degeneracy lifting between TO and LO branches are suggesting the importance of considering the NAC into the theoretical calculation. In order to calculate the phonon dispersion with nonanalytical term, the values of BEC for both of Sn and Te atoms are needed. So, the calculated value of BEC using the modern theory of polarization is found to be $\sim$7.5 ($\sim$-7.5) for Sn (Te) ion. In case of SnTe, we found from the literature that the value of high frequency static dielectric constant is 45\cite{dielectric1,dielectric2}. Therefore, using these dielectric constant and BEC, the phonon dispersion curve is calculated with including NAC and shown in Fig. 3(b). In case of acoustic phonon modes, it is seen from the figure that NAC does not show any significant change. But, the major effect on LO-TO band is found at $\Gamma$-point, where the LO-TO degeneracy is totally lifted with energy difference of $\sim$12 meV. However, this energy difference is not properly matching with the experimental observation, where it shows the difference $\sim$7.0 meV \cite{oneill}. Furthermore, it is observed that these phonon bands are strongly dependent on the values of BEC. Hence, it is evident that these calculated BEC are not appropriate. On fixing the value of BEC for Sn (Te) atom at $\sim$5.5 ($\sim$-5.5), the energy difference due to splitting of LO-TO branches is found to be $\sim$8.0 meV. These numbers provide the reasonable good matching with experiment\cite{oneill} in the range of experimental accuracy. The calculated phonon band structure using these values of BEC is shown in Fig. 3(c). The acoustic phonon branches obtained from this figure are nicely matched with the Fig. 3(a) and experimental acoustic phonon modes \cite{oneill}. The Fig. 3(c) shows that the maximum range of the acoustic phonon mode is reached at $\sim$12.5 meV, which is close to the estimated value by Pereira $et$ $al$ \cite{pereira}. Now, it is also seen from the figure that two TO branches are degenerate both at $\Gamma$ \& W-points and along $\Gamma$-X \& $\Gamma$-L directions. The major degeneracy lifting between these two optical phonon modes is found along X-W-$\Gamma$ and not seen near to W- and L-points. However, the top most phonon mode known as LO branch is totally non-degenerate with all other phonon modes along the observed k-direction. From the figure, the optical phonon band edge is seen at $\sim$15.3 meV, which is nicely agreed with the experimental data \cite{oneill,cowley}. Hence, it is noted that in order to study the phonon properties with better insight, it is necessary to consider the effect of NAC for this compound. So, it is important to note that all further calculations related to lattice vibration are performed using these phonon modes.

  The calculated phonon total DOS (TDOS) and PDOS of SnTe are shown in Fig. 3(d). The first peak's position corresponding to acoustic branches in phonon TDOS is found to be at $\sim$6 meV. Whereas, other two major peaks are observed at $\sim$10.7 meV and $\sim$12.0 meV, respectively. Additionally, one hump is also seen at $\sim$14.3 meV in phonon TDOS. In the inset of this figure, the calculated phonon TDOS is compared with the experimental data obtained from inelastic neutron scattering (INS) measurement by Li $et$ $al$ \cite{li}. It is found that all the peaks' position achieved from the first principle simulation are fairly good matched with the experimental phonon TDOS except the position of the hump, which is observed at $\sim$15.7 meV in experimental spectra. This may be due to the experimental broadening, which is not considered in our calculation. Now to observe the contribution to these peaks and hump from the different atoms, the phonon PDOS of Sn and Te atoms are discussed. It is clearly seen from the phonon PDOS that the first peak and the hump of phonon TDOS are mainly contributed by Sn atom. However, the other two main peaks ($i.e.$ at $\sim$10.5 meV and $\sim$12.0 meV) in phonon TDOS are formed due to a major contribution from Te atom. In case of peaks' and hump positions of phonon PDOS, it is important to note that the experimental observation using nuclear inelastic scattering (NIS) by Pereira $et$ $al$ \cite{pereira} shows nicely matching with our present work for Sn and Te atoms. Therefore, calculating different thermodynamic quantities of this compound using these phonon modes are expected to give good prediction.

\begin{figure}
  \begin{center}
    \includegraphics[width=0.82\linewidth, height=5.6cm]{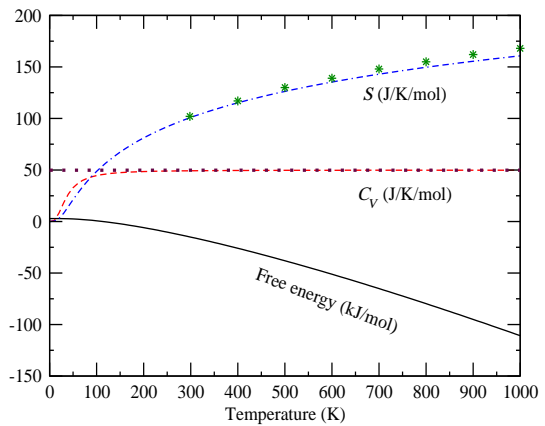} 
    \caption{(Colour online) Specific heat at constant volume ($C_V$), Helmholtz free energy ($F$) and entropy ($S$) for SnTe. Green star denotes experimental S values at different temperature\cite{pereira}.}
    \label{fig:}
  \end{center}
\end{figure} 
  
  Here, the calculated values of specific heat at constant volume ($C_V$), Helmholtz free energy ($F$) and entropy ($S$) for 0 K - 1000 K are shown in Fig. 4. The thermodynamic relations used to calculate all above mentioned quantities are given as \cite{togo},

 \begin{equation}
C_V={\sum\limits_{\textbf{q}j}}k_B\bigg(\frac{\hbar\omega_{\textbf{q}j}}{k_BT}\bigg)^2\frac{exp(\hbar\omega_{\textbf{q}j}/k_BT)}{[exp(\hbar\omega_{\textbf{q}j}/k_BT)-1]^2}
\end{equation}

\begin{equation}
 F=-k_BTlnZ
\end{equation}

where partition function ($Z$) represents as,

\begin{equation}
Z=exp(-\phi/k_BT)\prod_{\substack{\textbf{q}j}}\frac{exp(-\hbar\omega_{\textbf{q}j}/2k_BT)}{[1-exp(-\hbar\omega_{\textbf{q}j}/k_BT)]} \nonumber
\end{equation}

\begin{equation}
 S = -\frac{\partial F}{\partial T}
\end{equation}

where $\phi$ is the crystal potential energy, $k_B$ is the Boltzmann constant, $\hbar$ is reduced Planck's constant, $\omega_{\textbf{q}j}$ denotes the phonon frequency for mode of \textbf{q}, j and $T$ is temperature in absolute scale. In case of $C_V$, we know that in higher temperature, it obeys the Dulong-Petit law. In present case, the estimated value of $C_V$ from this law is $\sim$49.8 J/K/mol, which is represented by the dotted line in the figure. It is observed from this figure that the value of $C_V$ is found to reach the Dulong-Petit limit at $\sim$300 K, where it does not vary with increasing the temperature. However, in lower temperature, the $C_V$ is proportional to the $T^3$ as obvious from the Debey model. Moreover, the idea of Born-Oppenheimer approximation, which basically tells us to consider lattice as static, gives to reduce the complexity of calculation for finding the ground state energy eigenvalues through the DFT \cite{hao}. But for ions in solid, it is expected to have some vibrational energy even at zero temperature due to its quantum behaviour \cite{hao}. This so called zero-point energy can be obtained for any compound from the zero temperature value of $F$ \cite{togo,aschroft}. Thus, the calculated value of zero-point energy for SnTe obtained from $F$ is $\sim$2.8 kJ/mol. The figure shows that the temperature dependent $F$ value is decreasing monotonically upto 1000 K. This value becomes negative after $\sim$120 K. Now in case of $S$, the figure shows that the present computed values at different temperature is in good agreement with the estimation by Pashinkin $et$ $al$ \cite{pashinkin} in the temperature region 298 K - 1000 K. Another important thermodynamic quantity for any solid compound is the Debye temperature ($\Theta_D$). This $\Theta_D$ is such a transition point above where all the phonon modes are gained sufficient thermal energy to excite \cite{aschroft}. Nowadays, a plenty of different ways to find the value of $\Theta_D$, one of them is to consider the maximum phonon frequency ($\omega_{max}$) as Debye frequency \cite{aschroft}. This $\omega_{max}$ is found to be $\sim$16.3 meV obtained from the phonon TDOS in Fig. 3(d). Therefore, the calculated value of $\Theta_D$ for SnTe is $\sim$189 K, which is close to the estimated value by experimental measurement \cite{pereira}.

\subsection{Finite temperature electronic structure} 

\begin{figure}[]
   \begin{subfigure}{0.95\linewidth}
   \includegraphics[width=0.9\linewidth, height=6.0cm]{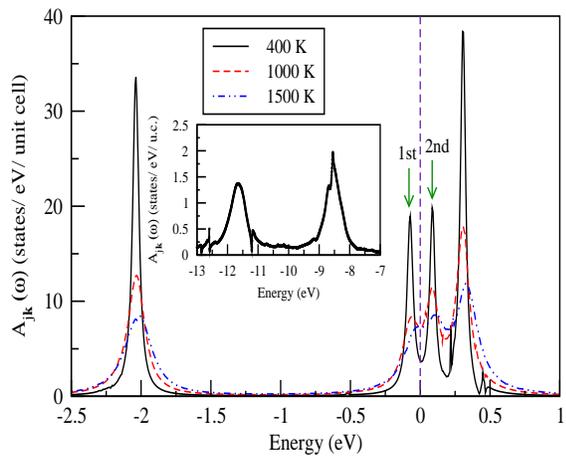}
   \caption{}
   \label{fig:} 
\end{subfigure}
\begin{subfigure}{0.95\linewidth}
   \includegraphics[width=0.9\linewidth, height=6.0cm]{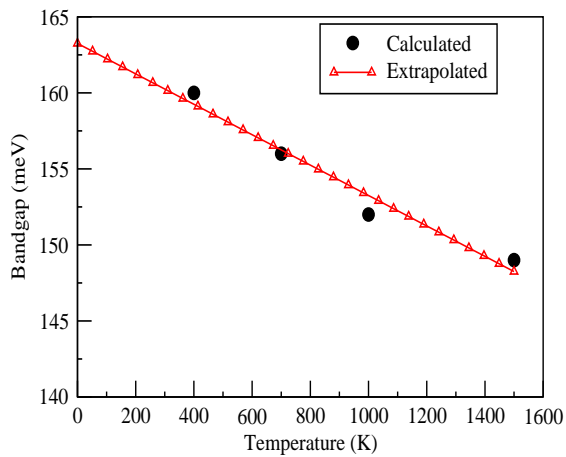}
   \caption{}
   \label{fig:}
\end{subfigure}
\begin{subfigure}{0.95\linewidth}
   \includegraphics[width=0.9\linewidth, height=6.0cm]{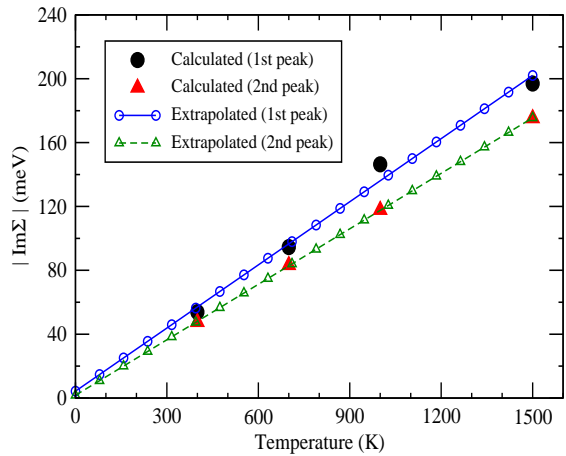}
   \caption{}
   \label{fig:}
\end{subfigure}
\caption{(a) Spectral functions of L-point from full-$GW$ at temperature 400 K (black solid line), 1000 K (red dashed line) and 1500 K (blue dotted dash line), (b) calculated (black dot) and extrapolated (red sold line) temperature dependent bandgap at L-point, (c) calculated (symbol) and extrapolated (line with symbol) $\arrowvert Im\Sigma (\omega)\arrowvert $ as function of temperature for 1st and 2nd peak.}
\end{figure}

  To get the insight of many-body interaction effect, a detailed temperature dependent electronic structure calculation using full-$GW$ method for this compound is discussed in this subsection. The spectral function ($A_{j\textbf{k}}(\omega)$) for the $j^{th}$ band at $\textbf{k}$-point is defined as\cite{chu-elk},
  
\begin{equation}
A_{j\textbf{k}}(\omega)=-\frac{1}{\pi}Im\,[G_{j\textbf{k}}(\omega)]
\end{equation}

where, $G_{j\textbf{k}}(\omega)$ is the Green function of interacting system. In order to compare the effect of electronic temperature on L-point, the $A_{j\textbf{k}}(\omega)$ at this point is calculated for 400 K, 1000 K and 1500 K, which are shown in Fig. 5(a). Here, we focused only on L-point because both the band inversion and bandgap are obtained near this high-symmetric \textbf{k}-point. It is clearly observed from the figure that the peaks' height (broadening) are decreasing (increasing) when the electronic temperature moves from 400 K to 1500 K. This behaviour follows as per the expected temperature dependence. The peak's center, which is marked as 1st (2nd), is considered as the quasiparticles' VB maximum (VBM) (CB minimum (CBM)). Therefore, the energy difference between 1st and 2nd peak's center is defined as the quasiparticle bandgap at this certain electronic temperature. However, we observed that above 1500 K, these two peaks are not well separated, which may be the effect of high temperature electronic excitation. Thus, the bandgaps are calculated upto 1500 K and extrapolated towards the lower temperature till $T$=0 K, which are provided in Fig. 5(b). The computed values of bandgap at 400 K, 700 K, 1000 K and 1500 K are $\sim$160 meV, $\sim$156 meV, $\sim$152 meV and $\sim$149 meV, respectively. It is observed from the figure that the values of bandgap are monotonically decreasing with increasing the temperature, which have a good agreement with previous theoretical and experimental works on this compound \cite{tsang,enomoto}. Recent theoretical work on this compound with considering the EPI is also suggesting the similar decreasing nature of bandgap with increasing temperature \cite{flores}. From our calculation, the decreasing rate of bandgap with respect to temperature upto 1500 K is found to be in order of 10$^{-5}$, which shows 1/10 times of the previous observation on this compound \cite{esaki,tsang}. This observed difference in rate of change of bandgap is attributed to only considering the EEI. But, the EPI and electron-defects interactions also play important role for obtaining the better bandgap, which are not included in our calculation. The extrapolated bandgap value at 300 K obtained after linear fitting of the calculated data is found to be $\sim$161 meV. This value shows only $\sim$19 meV difference from previous experimentally observed value at this temperature \cite{dimmock,esaki}. The value of bandgap at 0 K is found to be $\sim$163 meV from this extrapolated data, which is $\sim$38 ($\sim$37) meV lower than the calculated value obtained from $G_0W_0$ (PBEsol) method after including SOC. The SOC included full-$GW$ calculation is highly expensive in lower temperature. Thus, in present work, this calculation is not performed with SOC. However, the better insight of this electronic excitation can be found from the self-energy ($\Sigma (\textbf{k},\omega)$).

  It is known that the $\Sigma (\textbf{k},\omega)$ provides the direct information about the EEI. In general, $\Sigma (\textbf{k},\omega)$ is a complex quantity, in which real part of self energy $Re \Sigma (\textbf{k},\omega)$ can give information about the coherent weight of the spectrum. This coherent weight in the spectrum, which is also known as quasiparticle weight, is given by mass renormalization factor, $Z_\textbf{k}(\omega)$. The $Z_\textbf{k}(\omega)$ is defined as \cite{imada_rev},

\begin{equation}  
Z_\textbf{k}(\omega) = \Big[1-\frac{\partial \,Re \Sigma(\textbf{k},\omega)}{\partial(\omega)}\Big]^{-1}  
\end{equation}

However, \big(1 - $Z_\textbf{k}(\omega)$\big) provides information about the incoherent weight present in the spectrum. The calculated values of $Z_\textbf{k}(\omega)$ for 1st (2nd) peak at 400 K, 700 K and 1500 K are found to be $\sim$0.78 ($\sim$0.71), $\sim$0.76 ($\sim$0.73) and $\sim$0.79 ($\sim$0.75), respectively, which are not showing much difference with changing the electronic temperature. The value of $Z_\textbf{k}(\omega)$ obtained from $G_0W_0$ calculation at 0 K is $\sim$0.74 ($\sim$0.72). The effective mass of quasiparticle ($m^*$), which is equal to $\big(Z_\textbf{k}(\omega)\big)^{-1}$, of 1st (2nd) peak is found to be $\sim$1.3 ($\sim$1.4). At this point, it is important to note that $\sim$0.23 ($\sim$0.27) amount of spectral weight transfer occurs from coherent part to incoherent part of the spectrum for 1st (2nd) peak, which is observed from full-$GW$ calculation. This incoherent weight transfer is responsible for getting different satellite features in spectrum, which are observed at $\sim$-8.5 eV and shown in the inset of Fig. 5(a). Moreover, the another satellite peak is also found at $\sim$-11.6 eV as shown in the inset. The distance between the satellite peak observed at $\sim$-8.5 eV and the middle point of the energy $\sim$-0.25 eV to $\sim$0.5 eV is measured to be $\sim$8.6 eV. Similarly, it is found that the satellite peak observed at $\sim$-11.6 eV is kept distance by amount of $\sim$9.6 eV from the middle point of energy $\sim$-2.25 eV to $\sim$-1.75 eV. Because of the closeness of this distances, we may say that these two satellite peaks are obtained due to plasmon excitation and the plasmon frequency of this material is expected to be closer to $\sim$9.0 eV. At this point, it is important to note that the observation of negative Coulomb interaction within the $\omega$ of $\sim$8.5 to $\sim$14.3 eV in Fig. 2 may be related with this plasmon frequency as explained in subsection A.

  Now, in order to observe the lifetime of quasiparticle states, the temperature dependent $\arrowvert Im\Sigma (\omega)\arrowvert $ is calculated only for 1st and 2nd peaks which are shown in Fig. 5(c). It is known that the $\arrowvert Im\Sigma (\omega)\arrowvert $ is inversely related with the lifetime/relaxation time ($\tau $) of quasiparticle states. In this figure, a monotonically increasing nature for $\arrowvert Im\Sigma (\omega)\arrowvert$ is observed with varying the electronic temperature from 0 - 1500 K for both peaks. Thus, it is suggesting that $\tau $ is decreasing with increasing temperature. The calculated values of $\arrowvert Im\Sigma (\omega)\arrowvert $ for 400 K, 700 K, 1000 K and 1500 K of 1st (2nd) peak are $\sim$54 ($\sim$48) meV, $\sim$95 ($\sim$83) meV, $\sim$146 ($\sim$118) meV and $\sim$197 ($\sim$175) meV, respectively. The estimated value of slope after linear fitting of these 1st (2nd) peak's data is $\sim$0.132 ($\sim$0.115) meV/K. The extrapolated value of 1st (2nd) peak at 0 K is found to be $\sim$4 ($\sim$2) meV, which is nicely agreed with the calculated value obtained from ground state $G_0W_0$ method. Recently, Querales-Flores $et$ $al.$ have shown the temperature dependent $Im\Sigma (\omega)$ which is obtained by considering the EPI \cite{flores}. By comparing this with our present work, it suggests that in lower temperature the lifetime of electron due to EPI and EEI are comparable but not in case of higher temperature. Hence at this moment, it is noted that in higher temperature, the EEI are more important than EPI in understanding the transport behaviour of this compound. It is also proposed that these calculated results can be verified from the angle-resolved photoemission spectroscopy (ARPES) measurement. These results also clearly suggest the importance of EEI in understanding the physical properties of this compound. In the light of this, it would be interesting to calculate electronic lifetime due to EEI along the high-symmetric direction, which can also be verified from ARPES.

\begin{figure}[]
   \begin{subfigure}{0.96\linewidth}
   \includegraphics[width=0.92\linewidth, height=7.4cm]{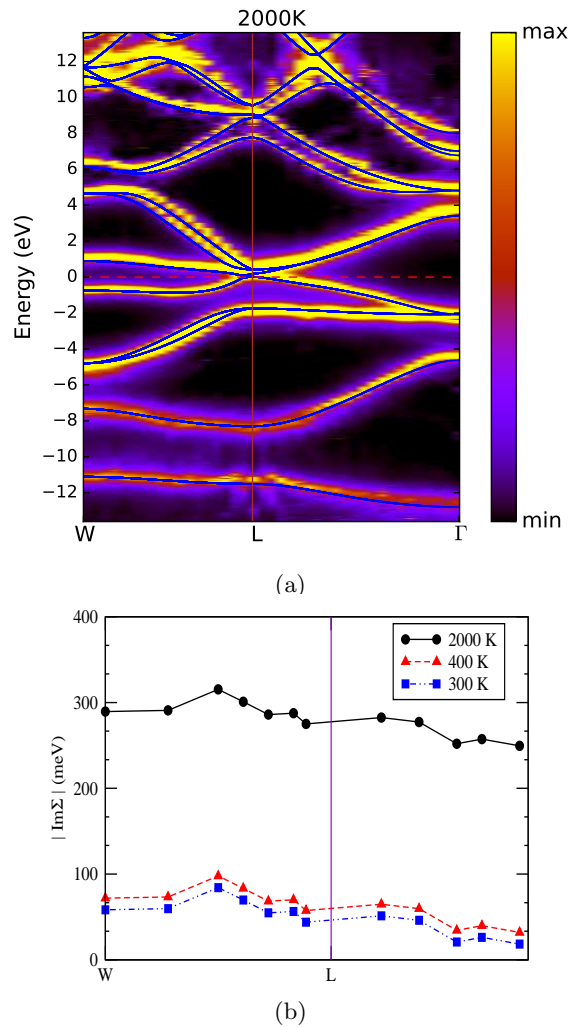}
   \caption{}
   \label{fig:} 
\end{subfigure}
\begin{subfigure}{0.9\linewidth}
   \includegraphics[width=0.82\linewidth, height=5.0cm]{fig6b.eps}
   \caption{}
   \label{fig:}
\end{subfigure}
\caption{(a) Momentum-resolved spectral function at 2000 K. Blue solid lines denote the DFT bands. (b) Calculated value of $\arrowvert Im\Sigma (\omega)\arrowvert $ along W-L-$\Gamma$ direction at 2000 K, 400 K and 300 K, respectively, for the top most VB.}
\end{figure}
  
  The calculated momentum-resolved spectral function of this compound using full-$GW$ method along the high symmetric k-direction of W-L-$\Gamma$ are shown in Fig. 6(a) at electronic temperature 2000 K. This calculation has high computational cost in lower electronic temperature. Therefore, comparatively higher electronic temperature ($i.e.$ 2000 K) is taken for this calculation without including SOC because it is observed that the $A_{j\textbf{k}}(\omega)$ is not significantly changed with SOC. In general, this $A_{j\textbf{k}}(\omega)$ have the information of both the coherent and incoherent part of the spectrum. Here in the figure, the yellowish (violetish blue) colour represents the coherent (incoherent) part of the spectrum. Dispersion curve obtained at DFT level using PBEsol XC functional is also given in this figure for comparison purpose. As per the expectation, more populated incoherent spectral weight is observed from the figure at this high temperature for both VB and CB. In order to discuss the EEI for the top most VB along W-L-$\Gamma$ k-direction, the calculated values of $\arrowvert Im\Sigma (\omega)\arrowvert $ of this band at 2000 K is plotted in Fig. 6(b). The value of $\arrowvert Im\Sigma (\omega)\arrowvert $ at W-point obtained from the figure is found to be $\sim$290 meV. This value increases along W-L direction and reaches at $\sim$315 meV around the middle of W- and L-points. After this point, the value is decreasing and showing $\sim$275 meV around L-point. Next along L- to $\Gamma$-point, the calculated value of $\arrowvert Im\Sigma (\omega)\arrowvert $ is changing slowly. After reaching the middle of L-$\Gamma$ direction, the minimum value of $\arrowvert Im\Sigma (\omega)\arrowvert $ is found to be $\sim$250 meV. But, it is very difficult to reach at such high temperature for the experimental observation by ARPES. In order to calculate the $\arrowvert Im\Sigma (\omega)\arrowvert $ at experimentally reachable temperature along W-L-$\Gamma$ direction, firstly we estimated the data for 400 K, which is shown in Fig. 6(b). It is observed from Fig. 5(c) for L-point that $\arrowvert Im\Sigma (\omega)\arrowvert $ is changing linearly with temperature ($i.e.$ $\arrowvert Im\Sigma (\omega)\arrowvert $ = (m\texttimes T) + c, where m is the slope and c is the value of $ \arrowvert Im\Sigma (\omega)\arrowvert $ at T = 0 K). In this case, the calculated value of m is $\sim$0.132 ($\sim$0.115) meV/K for 1st (2nd) peak. In addition to this, we already defined the 1st peak as the VBM, which is providing the information about a point of the top most VB. It is also observed that the values of $\arrowvert Im\Sigma (\omega)\arrowvert $ at W-point, which is corresponding to top most VB's peak in $A_{j\textbf{k}}(\omega)$, are also showing linear behaviour with changing the temperature. The value of m is estimated to be $\sim$0.14 meV/K for this peak. It is noted that this point also belongs to one of a point of the top most VB. Thus, considering all these aspect, if we assume that the values of $\arrowvert Im\Sigma (\omega)\arrowvert $ along W-L-$\Gamma$ direction will also show the linear temperature dependence. Then, under this assumption, the averaged value of m \big($i.e$ average value of m = (value of m at W-point + value of m at L-point) / 2\big) is taken as 0.136 meV/K for calculating the $\arrowvert Im\Sigma (\omega)\arrowvert $ along this k-direction for 400 K by using the data of 2000 K. Particularly this lower temperature ($i.e.$ 400 K) is chosen because the value of $\arrowvert Im\Sigma (\omega)\arrowvert $ for W- and L-points obtained from the above mentioned procedure can be possible to compare with the calculated value from $A_{j\textbf{k}}(\omega)$. Thus at the W (L)-point, the amount of variation obtained from these two different procedure of finding the $\arrowvert Im\Sigma (\omega)\arrowvert $ is found to be $\sim$20 ($\sim$6) meV. Thus, this procedure appears reasonably good for estimating the values of $\arrowvert Im\Sigma (\omega)\arrowvert$ along this high-symmetric k-direction for other lower temperature. In overall along W-L-$\Gamma$ direction, the difference between the data of $\arrowvert Im\Sigma (\omega)\arrowvert$ for 2000 K and 400 K is $\sim$217.6 meV. The maximum (minimum) value of the $\arrowvert Im\Sigma (\omega)\arrowvert$ at 400 K is found to be $\sim$98 ($\sim$33) meV around the middle of W-L (L-$\Gamma$) direction. Now, the same procedure is applied to calculate the values of $ \arrowvert Im\Sigma (\omega)\arrowvert $ along this k-direction for 300 K, which is the most easily accessible temperature for experimental measurement. These calculated values are also plotted in Fig. 6(b). The calculated value of $ \arrowvert Im\Sigma (\omega)\arrowvert $ at W (L)-point is found to be $\sim$58 ($\sim$46) meV for 300 K. At this temperature, the maximum and minimum values are estimated to be $\sim$84 meV and $\sim$19 meV, respectively, at the similar position as mentioned in 400 K. The overall difference of $\arrowvert Im\Sigma (\omega)\arrowvert$ along this high symmetric k-direction between 2000 K (400 K) and 300 K is estimated to be $\sim$231.2 ($\sim$13.6) meV. At this point, it is mentioned that this compound contains surface state and also classified as a non-trivial TCI. Hence, it is also interesting to study the surface properties of this compound which is discussed in next subsection.

\subsection{Robustness of (001) surface states under defects}

\begin{figure}[]
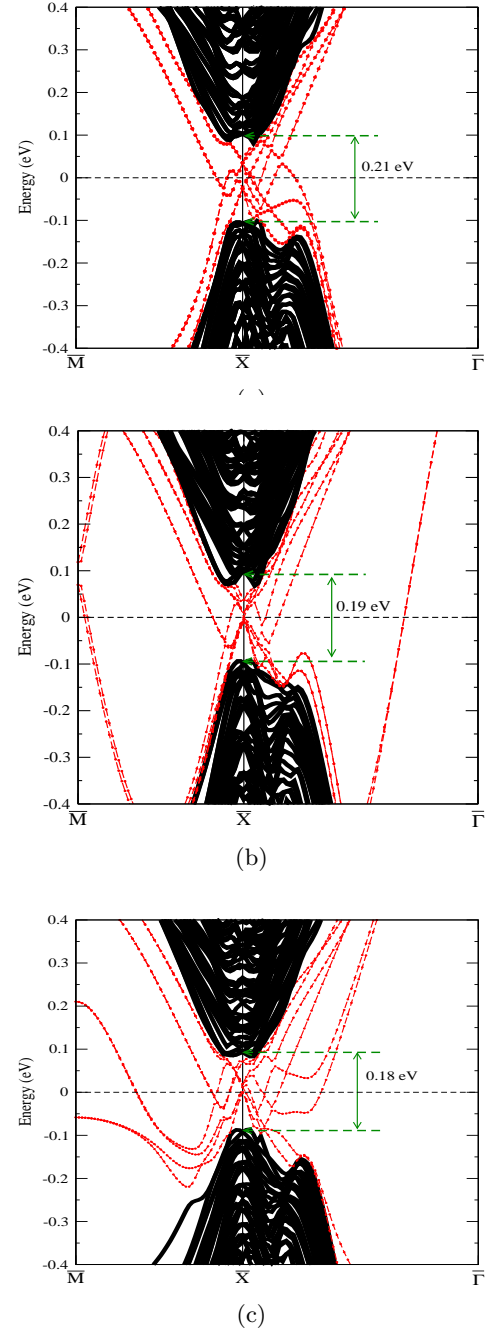

   \begin{subfigure}{0.85\linewidth}
   \includegraphics[width=0.85\linewidth, height=5.4cm]{fig7a.eps}
   \caption{}
   \label{fig:} 
\end{subfigure}
\begin{subfigure}{0.85\linewidth}
   \includegraphics[width=0.85\linewidth, height=5.4cm]{fig7b.eps}
   \caption{}
   \label{fig:}
\end{subfigure}
\begin{subfigure}{0.85\linewidth}
   \includegraphics[width=0.85\linewidth, height=5.4cm]{fig7c.eps}
   \caption{}
   \label{fig:}
\end{subfigure}
\caption{Surface band structure of (001) surface of 14 layers for (a) pure SnTe, (b) point defect due to Te vacancy (c) line defect due to Sn and Te vacancies.}
\end{figure}

  In general, it is known that in topological materials, metallic surface states are present. In case of SnTe, these surface states are found both for (001) and (111) surfaces \cite{liu}. For (001) surface, both theoretical\cite{hsieh} and experimental\cite{tanaka} studies reveal the presence of Dirac point (DP) at $E_F$ along $\bar{\Gamma}$-$\bar{X}$ direction in 2D brillouin zone (BZ). Moreover, these surface states are not protected by TRS, whereas they are protected by mirror symmetry. Liu $et$ $al.$ predicts from their theoretical observation that this surface is capable for using as a topological transistor device \cite{jliu2014}. It is also known that the DP of this surface can be moved towards the $\bar{X}$ ($\bar{\Gamma}$)-point due to tensile (compressive) strain \cite{zeljkovic}. Thus, if the DP is obtained at $\bar{X}$-point or $\bar{\Gamma}$-point, then it will be protected both by mirror symmetry and TRS. Therefore, this will enable more option to design several devices via playing with both symmetries. In order to explore the possibility of this aspect, the unrelaxed supercell calculation with same lattice parameter and fractional position of bulk unit cell of this compound is carried out for different number of layers. The definition of the layer is already given in section II. After observing the surface states for different number of layers, we found that 14 layers calculation gives the DP at $\bar{X}$-point. The dispersion curve obtained from this calculation is shown in Fig. 7(a), where the surface (bulk) bands are coloured by red dashed (black solid). These bands are called as surface bands, because they are not seen in bulk band structure of this compound. These surface states are showing the linear band dispersion relation at this DP, which is as per the expectation. This figure clearly shows the presence of four-fold degeneracy of this DP, which is protected by both of mirror and TRS. It is known that the projection of L-point of 3D face-centered-cubic (FCC) BZ on 2D BZ of (001) surface gives the $\bar{X}$-point. Thus, the observed bandgap of $\sim$0.21 eV at $\bar{X}$-point is found to be less than the calculated bulk value of $\sim$0.30 eV at L-point from PBEsol calculation. In the figure, the zero energy is fixed at the middle of the bulk bandgap.  At this point, it is important to note that the internal strain formed due to this structure may be responsible for getting the DP at $\bar{X}$ for (001) surface. It is suggested that this theoretically simulated structure can be possible to prepare using Molecular-beam epitaxy (MBE). ARPES can also verify the existence of this DP at $\bar{X}$-point after making this sample using MBE. However, it is known that in practical situation, every compound has some defects. In order to observe the robustness of these surface bands, the same calculations are carried out after creating different type of defects in the pure structure.

   Recently, the experimental observation on (001) surface showed the proof of Sn vacancies, which is also verified by theoretical simulation \cite{dzhang}. In addition to this, Zhang \textit{et} \textit{al.} suggest the presence of many other defects for this surface, which are observed from the scanning tunneling microscopy (STM) image \cite{dzhang}. But, they did not discuss about these defects in details. Here, motivated from this, some common defects ($e.g.$ point and line defects) on (001) surface are considered to study, where the pure surface is obtained with same procedure as mentioned above. The point defect due to Te vacancy is created by removing one top most Te ion from the 14$^{th}$ layer of pure supercell structure. The dispersion curve obtained from this calculation is shown in Fig. 7(b). It is observed from the figure that the DP is positioned at $\bar{X}$-point. This DP is found to be two-fold degenerate. This point defect breaks the mirror symmetry in this crystal structure. Therefore, the DP is only protected by TRS. In this case, the estimated value of bulk bandgap is $\sim$0.19 eV at $\bar{X}$-point. It is also noted that the DP is found within this bulk bandgap, where the zero energy is set to be at the middle of this bandgap. Now similarly, for creating the line defect in pure structure, we removed two ions, which are the top most Sn and Te ions in 14$^{th}$ layer of pure supercell structure. These ions are seating consecutively at the fractional positions of this crystal structure. Here, we considered only those two ions which are not seated on the mirror plane of (1-10). The calculated dispersion curve after considering this line defect is plotted in Fig. 7(c). It is seen from the figure that two-fold degenerate DP is found at $\bar{X}$-point. In this scenario, the calculated bulk bandgap is found to be $\sim$0.18 eV. In this case also zero energy is put at the middle of this bandgap. This line defect helps to break the mirror symmetry for this crystal structure. Thus in this case, TRS is only protecting the DP. Therefore, it is predicted that the metallic surface bands for (001) surface are showing their existence even in the presence of these two types of defects. Hence, it provides the practical usefulness of this surface for different device application due to the presence of robust metallic surface state.

\section{Conclusions} 

  In this work, we have investigated the detailed temperature dependent electronic structure and thermodynamic properties of SnTe by using $state$ $of$ $the$ $art$ DFT and $GW$ calculations. The calculated bandgaps from PBEsol and $G_0W_0$ are in good agreement with the experimental data, while mBJ is providing quite lower bandgap value than experiment suggesting further improvement of this method. The computed value of fully screened Coulomb interaction ($W$) for Sn (Te) 5$p$ orbitals is found to be $\sim$1.39 ($\sim$1.70) eV. It is seen by observing the nature of $W(\omega)$ that full-$GW$ method can sufficiently describe the excited electronic state of this compound. The deeper understanding of this $W(\omega)$ can be obtained with the discussion of plasmon excitation. In order to explain the experimental phonon modes, long range Coulomb forces through nonanalytical term correction are need to be included in calculation. The temperature dependent electronic structure study provides the decreasing behaviour of bandgaps with increasing temperature. The estimated value of bandgap at 300 K is found to be $\sim$161 meV. Present study reveals the importance of electron-electron interactions to study the high temperature transport behaviour of this compound. The $\arrowvert Im\Sigma (\omega)\arrowvert$ along W-L-$\Gamma$ direction are estimated for 300 K, which can be verified from ARPES. In case of (001) surface, the Dirac cone is found at $\bar{X}$-point from 14 layers calculation. The surface states are robust in presence of different types of defects. Thus, this compound may become a potential candidate for device making application.

\end{document}